\documentclass[preprint,prd,eqsecnum,aps,epsf]{revtex4}

\usepackage{axodraw}

\begin{document}

\def\aprge{\buildrel > \over {_{\sim}}}
\def\aprle{\buildrel < \over {_{\sim}}}

\def\etal{{\it et.~al.}}
\def\ie{{\it i.e.}}
\def\eg{{\it e.g.}}

\def\bwt{\begin{widetext}}
\def\ewt{\end{widetext}}
\def\be{\begin{equation}}
\def\ee{\end{equation}}
\def\bea{\begin{eqnarray}}
\def\eea{\end{eqnarray}}
\def\bean{\begin{eqnarray*}}
\def\eean{\end{eqnarray*}}
\def\bary{\begin{array}}
\def\eary{\end{array}}
\def\bi{\bibitem}
\def\bit{\begin{itemize}}
\def\eit{\end{itemize}}

\def\lan{\langle}
\def\ran{\rangle}
\def\lra{\leftrightarrow}
\def\la{\leftarrow}
\def\ra{\rightarrow}
\def\dash{\mbox{-}}
\def\ol{\overline}

\def\ub{\ol{u}}
\def\db{\ol{d}}
\def\sb{\ol{s}}
\def\cb{\ol{c}}

\def\re{\rm Re}
\def\im{\rm Im}

\def \b{{\cal B}}
\def \ca{{\cal A}}
\def \ko{K^0}
\def \ok{\overline{K}^0}
\def \s{\sqrt{2}}
\def \st{\sqrt{3}}
\def \sx{\sqrt{6}}
\title{\Large{\bf Spin Polarisability of the Nucleon \\
in the Heavy Baryon Effective Field Theory}}
\author{$^*$K.B. Vijaya Kumar, Yong-Liang Ma and Yue-Liang Wu}
\address{Institute of Theoretical Physics, Chinese Academy of Sciences, Beijing
100080, China}\footnotetext{$^*$ On leave from Department of
Physics, University of Mangalore, Mangalore 574 199,
India.}
\date{\today}
\begin{abstract}
We have constructed a heavy baryon effective field theory with
photon as an external field in accordance with the symmetry
requirements similar to the heavy quark effective field theory. By
treating the heavy baryon and anti-baryon equally on the same
footing in the effective field theory, we have calculated the spin
polarisabilities $\gamma_i, i=1\cdots4$ of the nucleon at third
order and at fourth-order of the spin-dependent Compton
scattering. At leading order (LO), our results agree with the
corresponding results of the heavy baryon chiral perturbation
theory, at the next-to-leading order(NLO) the results show a large
correction to the ones in the heavy baryon chiral perturbation
theory due to baryon-antibaryon coupling terms. The low energy
theorem is satisfied both at LO and at NLO. The contributions
arising from the heavy baryon-antibaryon vertex were found to be
significant and the results of the polarisabilities obtained from
our theory is much closer to the experimental data.

\end{abstract}
\maketitle

\section{Introduction}
The Compton-scattering process is an important tool to probe the
nucleon structure by measurments of various polarisabilities. For
unpolarised proton the experimental amplitude is well determined
and is in good agreement with the results of heavy-baryon chiral
perturbation theory (HBCHPT). But, with regard to the scattering
from polarised targets, it is less satisfactory.  At present,
there exists no direct measurments of the polarisabilities of
polarised Compton scattering. For the spin dependent pieces  the
scattering amplitude in the Breit frame is:
\begin{eqnarray}
T&=&\epsilon^{\prime\mu}\Theta_{\mu\nu}\epsilon^\nu\nonumber\\
&=&i\vec{\sigma}\cdot(\vec{\epsilon}^{~\prime}\times\vec{\epsilon})A_3(\omega,\theta)+i\vec{\sigma}\cdot(\hat{\vec{k}}^{\prime}\times\hat{\vec{k}})\vec{\epsilon}^{~\prime}\cdot\vec{\epsilon}A_4(\omega,\theta)\nonumber\\
&&+[i\vec{\sigma}\cdot(\vec{\epsilon}^{~\prime}\times\hat{\vec{k}})\vec{\epsilon}\cdot\hat{\vec{k}}^{\prime}-i\vec{\sigma}\cdot(\vec{\epsilon}\times\hat{\vec{k}}^{\prime})\vec{\epsilon}^{~\prime}\cdot\hat{\vec{k}}]A_5(\omega,\theta)\nonumber\\
&&+[i\vec{\sigma}\cdot(\vec{\epsilon}^{~\prime}\times\hat{\vec{k}}^{\prime})\vec{\epsilon}\cdot\hat{\vec{k}}^{\prime}-i\vec{\sigma}\cdot(\vec{\epsilon}\times\hat{\vec{k}})\vec{\epsilon}^{~\prime}\cdot\hat{\vec{k}}]A_6(\omega,\theta)\nonumber\\
&&+i\vec{\sigma}\cdot(\hat{\vec{k}}^{\prime}\times\hat{\vec{k}})\vec{\epsilon}^{~\prime}\cdot\hat{\vec{k}}\vec{\epsilon}\cdot\hat{\vec{k}}^{\prime}A_7(\omega,\theta)+\mbox{spin
independent terms}.\label{amplitude1}
\end{eqnarray}
where $\omega$ is the  photon energy and $k$ is the incoming
photon momentum, $\epsilon$ and $\epsilon^\prime$  are the
incoming and outgoing  photon polarization directions respectively
and the hats indicate unit vectors. By crossing symmetry the
functions $A_{i}$ are odd in $\omega$. The leading pieces in the
expansion are governed by low-energy theorems\cite{Op16}, and the
next order terms contain the spin polarisabilities $\gamma_i$
\begin{eqnarray}
A_3(\omega,\theta)&=&\frac{e^2\omega}{2m_B}[Q(Q+2\kappa)-(Q+\kappa)^2\cos\theta]+4\pi\omega^3(\gamma_1+\gamma_5\cos\theta)\nonumber\\
&&-\frac{e^2Q(Q+2\kappa)\omega^3}{8m^4_B}+{\cal
O}(\omega^5)\nonumber\\
A_4(\omega,\theta)&=&-\frac{e^2\omega}{2m^2_B}(Q+\kappa)^2+4\pi\omega^3\gamma_2+{\cal
O}(\omega^5)\nonumber\\
A_5(\omega,\theta)&=&\frac{e^2\omega}{2m^2_B}(Q+\kappa)^2+4\pi\omega^3\gamma_4+{\cal
O}(\omega^5)\nonumber\\
A_6(\omega,\theta)&=&-\frac{e^2\omega}{2m^2_B}Q(Q+\kappa)+4\pi\omega^3\gamma_3+{\cal
O}(\omega^5)\nonumber\\
A_7(\omega,\theta)&=&{\cal O}(\omega^5)
\end{eqnarray}
where the charge of the nucleon is $Q=(1+\tau_3)/2$ ($\tau_3$ is
the third component of the isospin) and its anomalous magnetic
moment is $\kappa=(\kappa_s+\kappa_v\tau_3)/2$. Only four of the
polarisabilities are independent due to the relation
$\gamma_5+\gamma_2+2\gamma_4=0$.

At present there are only estimates of the polarisabilities. The
best estimate exists for forward scattering where only $A_3$
contributes. The quantity $4 \pi f_2(0)$ is defined as
$dA_3(\omega,0)/d\omega$ at $\omega=0$, and depends only on
$\kappa^2$\cite{Op16} according to the low energy theorem (LET).
The relevant polarisability is $\gamma_0=\gamma_1+\gamma_5$, which
is related via a dispersion relation to measurements at energies
above the threshold for pion photo production, $\omega_0$:
\begin{eqnarray}
\gamma_0&=&\frac{1}{4\pi^2}\int_{\omega_0}^\infty d\omega
\frac{\sigma_-(\omega)-\sigma_+(\omega)}{\omega^3}
\end{eqnarray}
where $\sigma_\pm$ are the parallel and antiparallel
cross-sections for photo absorption and the related sum rule for
the model-independent piece, due to Gerasimov, Drell and
Hearn\cite{Op17}, has the same form except that $1/\omega$
replaces  $1/\omega^3$.

Very recently, the various measurements have been made with MAMI
at Mainz\cite{Op18}, for photon energies between $200\sim800 MeV$;
the range will be extended downward to 140 MeV, and a future
experiment at Bonn will extend it upwards to 3 GeV. The MAMI data
does not currently go low enough in energy to give a reliable
result for the spin polarisability $\gamma_0$. However
electroproduction data have also been used to extract this
quantity; Sandorfi et al.\cite{Op19} find $ \gamma^p_{0}= -1.3
\times 10^{-4}  fm^4$ and $ \gamma^n_{0}= -0.4 \times 10^{-4}
fm^4$, while a more recent analysis of Drechsel et.al\cite{Op20}
gives a rather smaller value of $ \gamma^p_{0}= -0.6 \times
10^{-4} fm^4$. (We shall use units of $10^{-4} fm^4$ for
polarisabilities from now on). The spin polarisability has been
calculated in the frame work of HBCHPT : at lowest (third) order
in the expansion $\gamma_0=\alpha_{em}g^2_A/(24\pi^2f_\pi^2
m^2_\pi)=4.51$ which diverges as $1/m_\pi^2$ in the chiral limit.
Here, the entire contribution comes from $\pi N$ loops. Being a
spin dependent quantity, the spin polarisability should receive a
sizable contribution from the delta resonance. The effect of the
$\Delta$ enters in counter-terms at fifth order in standard
HBCHPT, and has been estimated to be so large as to change the
sign\cite{Op21}. The calculation has also been done in an
extension of HBCHPT with an explicit $\Delta$ by Hemmert et
al.\cite{Op22}. They find that the principal effect is from the
$\Delta$ pole, which contributes $-2.4$ with the effect of
$\pi\Delta$ loops being small $-0.2$. Also, one other combination
of the polarisabilities, the backward scattering,
$\gamma_\pi=\gamma_1-\gamma_5$ has been estimated from the
low-energy data for Compton scattering from the proton by Tonnison
et al.\cite{Op23}. The backward scattering is dominated by the
anomalous $\pi^0$ exchange graph, which vanishes for the forward
scattering, but at third order and at fourth order there are also
pion loop contributions. The experimental value is
$\gamma_\pi=-27.1$ with experimental and theoretical errors of
about $10\%$ each. The HBCHPT results of Hemmert et al. is $-36.7$
of which $-43.5$ is the anomalous contribution, $4.6$ is the $\pi
N$ piece and $2.2$ comes from including the $\Delta$ \cite{Op22}.

The fourth order contribution to all the four polarisabilities has
been worked by several groups in the frame work of HBCHPT
\cite{Op41}\cite{Op42}\cite{Op43}\cite{argue}. At $O(p^4)$ there
are no seagulls and since the NLO pieces of the $A_i(\omega)$ are
of the fourth chiral order and are odd in $\omega$, they will have
expansions of the form $e^2\omega(a m_\pi + b
\frac{\omega^3}{m_\pi}+...)$. These non-analytic powers of
$m^2_\pi$ cannot be present in the basic couplings in the
Lagrangian, but can only be generated from  loops. In the work of
Gellas et al.\cite{Op41} they have not included the one-particle
reducible graphs (Fig.2g in our paper) in their definition of the
polarsabilities. With the addition of that contribution their
results agree with the results of ref.\cite{Op42}\cite{Op43}. The
polarisability of interest, the forward spin polarisability to
order $O(p^4)$  turns out to be $\gamma_0=4.5-(6.9+ 1.5\tau_3)$,
where the first term in the above expression is the contribution
from the leading order. The NLO contributions are large and hence
call the convergence of the expansion into question.  As has been
argued by Bernard et.al \cite{ber2003}the bad reason for the
convergence is entirely related to the contributions from the Born
terms. It has been argued that the calculations of the Born graphs
to the fourth order is not sufficient to obtain convergence and
hence is necessary to take into account the two-loop corrections
which appears at the fifth order\cite{jam1999}.

It is to be noted that HBCHPT \cite{HBreview1} is an effective
theory constructed by using the idea of heavy quark effective
theory (HQET)\cite{HQET}. In HBCHPT, baryon is considered as
extremely heavy and only baryon momenta relative to the rest mass
will count which is very small. The heavy source (baryon) is
surrounded by a cloud of almost massless particles which is
exactly the idea used in HQET. In HQET, after decoupling the
"quark fields" and "antiquark fields" only one of them is treated
independently. Strictly speaking, in quantum field theory,
particle and antiparticle decouple completely only in the infinite
heavy quark mass limit $m_Q\rightarrow\infty$. To consider the
finite quark mass correction, it is necessary to include the
contribution from the components of the antiquark fields. For
that, one can simply extend the usual HQET \cite{HQET} to a heavy
quark effective field theory (HQEFT) with keeping both effective
quark and antiquark fields. This was first pointed out by one of
us \cite{HQEFT1}, where a new formulation of heavy quark effective
Lagrangian was derived from the full QCD. Its form permits an
expansion in powers of the heavy quark momentum characterizing its
off-shellness divided by its mass. A detailed comparison between
HQEFT and HQET is provided in the reference \cite{HQEFT2}. It is
important to note that at the leading order the HQEFT is same as
HQET,  the difference arises from the sub-leading terms which is
proportional to the inverse of heavy quark mass $m_Q$. The reason
is that in the construction of HQET, the particle and
anti-particle are independently treated based on the assumption
that the particle number and anti-particle number are conserved
separately in the effective theory. Such an assumption is valid
only in the infinite mass limit.  Hence, quark-anti-quark coupled
terms that correspond to the pair creation and annihilation
interaction terms in full QCD were inappropriately dropped away in
the usual HQET. Those terms have been shown in HQEFT of QCD to be
suppressed by $1/m_Q$ and they become vanishing in the infinite
mass limit. Thus, the usual HQET based on the assumption that the
quark and anti-quark number conservation in the effective
Lagrangian is an incomplete theory for evaluating the subleading
order corrections. Unlike the derivation of HQET from QCD by
making the assumption of particle and anti-particle conservation
in the effective theory, the HQEFT was derived from QCD by
treating all the contributions of the field components, i.e. large
and small, particle and anti-particle in the effective Lagrangian,
so that the resulting effective lagrangian forms the basis for a
complete effective field theory of heavy quarks. A more systematic
construction and detailed interpretation of the HQEFT is recently
presented in\cite{HQEFT3}. The HQEFT has been used to study heavy
quark systems \cite{W1,W2,W3,W4,W5,W6,W7}. Considering the
shortcomings of the HQET and the rationality of HQEFT from the
view point of quantum field theory, it is more reasonable to
construct the chiral theory of baryons in the frame work of HQEFT.
In a word, the HQET is not a complete theory from the view point
of quantum field theory, so that the HBCHPT which was constructed
by using the idea of HQET is also not complete. To construct a
complete heavy baryon effective field theory (HBEFT), a similar
idea of HQEFT should be used by treating heavy baryon and
antibaryon fields equally on the same footing.

The paper is organised as follows. In Sec.II, we will construct
the heavy baryon effective field theory (HBEFT) that contains both
effective baryon and antibaryon fields. We construct ${\cal
L}_{\pi N}^{(1)}$ and ${\cal L}_{\pi N}^{(2)}$ with photon as an
external field. The constructed Lagrangian has all the symmetries
of the HBCHPT. The Feynman rules from the leading order and next
to leading order Lagrangian required in the study of Compton
scattering are derived. In Sec.III, we calculate the spin
polarisabilities at the lowest order and at NLO. In Sec.IV, we
discuss the complete results of the fourth order calculations and
the important conclusions. Appendix A gives the pertinent Feynman
rules from ${\cal L}_{\pi N}^{(1)}$ and ${\cal L}_{\pi N}^{(2)}$.
Appendix B gives the full amplitude for the diagrams of Fig.2 and
Fig.3.

\section {Construction of the Effective Lagrangian ${\cal L}_{\pi N}^{(1)}$
and ${\cal L}_{\pi N}^{(2)}$}

Here, we briefly describe the new formulation of the will be
constructed HBEFT that contains both effective baryon and
anti-baryon fields. In chiral perturbation theory, the
transformation properties of baryon and meson fields under chiral
symmetry $SU(2)_L\times SU(2)_R$ are\cite{HBreview1}
\begin{eqnarray}
&&U(x)\rightarrow g_LU(x)g^\dag_R,~~~~g_L\times g_R\in
SU(2)_L\times
SU(2)_R\\
&&H(x)\rightarrow G(x)H(x),~~~~G(x)\in SU(2)_{local}
\end{eqnarray}
where $H(x)$ is baryon matrix
\begin{eqnarray}
H(x)=\left(%
\begin{array}{c}
  p \\
  n \\
\end{array}%
\right)
\end{eqnarray}
$G(x)$ is determined by
\begin{eqnarray}
\xi(x)\rightarrow g_L\xi(x)G^\dag(x)=G(x)\xi(x)g^\dag_R
\end{eqnarray}
and is hidden local symmetry\cite{hidden}. It can be verified that
$SU(2)_{local}=SU(2)_V$ when $g_L=g_R$. In the following, we
select
\begin{eqnarray}
U(x)=\xi^2(x)=e^{2i\Pi(x)\over f_\pi}
\end{eqnarray}
and $f_\pi$ is the pion decay constant. The covariant derivative
appearing in the  kinetic energy term of the baryon can be written
as
\begin{eqnarray}
&&D_\mu H(x)=\partial_\mu H(x)-iV_\mu H(x)\\
&&V_\mu(x)={i\over2}[\xi(x)\partial_\mu\xi^\dag(x)+\xi^\dag(x)\partial_\mu\xi(x)]
\end{eqnarray}
under chiral transformation, it transforms as
\begin{eqnarray}
D_\mu H(x)\rightarrow G(x)D_\mu H(x)
\end{eqnarray}

Besides the quantities introduced above, we have,
\begin{eqnarray}
A_\mu={i\over2}[\xi(x)\partial_\mu\xi^\dag(x)-\xi^\dag(x)\partial_\mu\xi(x)]
\end{eqnarray}
under chiral transformation it transforms as
\begin{eqnarray}
A_\mu\rightarrow G(x)A_\mu G^\dag(x)
\end{eqnarray}

Then the lowest order Lagrangian is
\begin{eqnarray}
{\cal L}&=&{\rm Tr}[\bar{H}(iD\hspace{-0.25cm}\slash-m_B)H]+2D{\rm
Tr}[\bar{H}A\hspace{-0.2cm}\slash\gamma_5H]\label{LFull}
\end{eqnarray}
where $m_B$ is the baryon mass matrix and the coupling constant
$D$ can be determined phenomenologically.

We expand the above Lagrangian in terms of the inverse of the
heavy baryon mass similar to the HQEFT.  We are making use of the
some of the conclusions given in\cite{HQEFT1,HQEFT2,HQEFT3}.

Baryon fields can be decomposed  into baryon and anti-baryon which
correspond to the positive and negative solutions respectively of
the Dirac equation, Using the relation
\begin{eqnarray}
&&H=\bigg[1+\bigg(1-\frac{iv\hspace{-0.2cm}\slash v\cdot
D+m_B}{2m_B}\bigg)^{-1}\frac{iD\hspace{-0.25cm}\slash_\bot}{2m_B}\bigg]\hat{H}_v\nonumber\\
&&\bar{H}=\bar{\hat{H}}_v\bigg[1+\frac{-i\overleftarrow{D\hspace{-0.25cm}\slash_\bot}}{2m_B}\bigg(1-\frac{-iv\hspace{-0.2cm}\slash
v\cdot \overleftarrow{D}+m_B}{2m_B}\bigg)^{-1}\bigg]
\end{eqnarray}
integrating  out the small components of baryon and anti-baryon
field in $(\ref{LFull})$ we get,
\begin{eqnarray}
{\cal L}_v&=&{\rm Tr}[\bar{\hat{H}}_v(iv\hspace{-0.2cm}\slash
v\cdot D-m_B)\hat{H}_v]+{1\over2m_B}{\rm
Tr}[\bar{\hat{H}}_viD\hspace{-0.25cm}\slash_\perp(1-\frac{iv\hspace{-0.2cm}\slash
v\cdot
D+m_B}{2m_B})^{-1}iD\hspace{-0.25cm}\slash_\perp\hat{H}_v]\nonumber\\
&&+{1\over2m_B}{\rm
Tr}[\bar{\hat{H}}_viD\hspace{-0.25cm}\slash_\perp(1-\frac{iv\hspace{-0.2cm}\slash
v\cdot D+m_B}{2m_B})^{-1}(iv\hspace{-0.2cm}\slash v\cdot
D-m_B)\hat{H}_v]\nonumber\\
&&+{1\over4m^2_H}{\rm
Tr}[\bar{\hat{H}}_v(-i\overleftarrow{D\hspace{-0.25cm}\slash_\perp})\bigg(1-\frac{-iv\hspace{-0.2cm}\slash
v\cdot
\overleftarrow{D}+m_B}{2m_B}\bigg)^{-1}iD\hspace{-0.25cm}\slash_\perp\bigg(1-\frac{iv\hspace{-0.2cm}\slash
v\cdot
D+m_B}{2m_B}\bigg)^{-1}(iD\hspace{-0.25cm}\slash_\perp)\hat{H}_v]\nonumber\\
&&+2D{\rm
Tr}\bigg\{\bar{\hat{H}}_v\gamma_\mu\gamma_5A_\mu\hat{H}_v+\bar{\hat{H}}_v\gamma_\mu\gamma_5A_\mu\bigg(1-\frac{iv\hspace{-0.2cm}\slash
v\cdot
D+m_B}{2m_B}\bigg)^{-1}\frac{iD\hspace{-0.25cm}\slash_\bot}{2m_B}\hat{H}_v\nonumber\\
&&+\bar{\hat{H}}_v\frac{-i\overleftarrow{D\hspace{-0.25cm}\slash_\bot}}{2m_B}\bigg(1-\frac{-iv\hspace{-0.2cm}\slash
v\cdot
\overleftarrow{D}+m_B}{2m_B}\bigg)^{-1}\gamma_\mu\gamma_5A_\mu\hat{H}_v\nonumber\\
&&+\bar{\hat{H}}_v\frac{-i\overleftarrow{D\hspace{-0.25cm}\slash_\bot}}{2m_B}\bigg(1-\frac{-iv\hspace{-0.2cm}\slash
v\cdot
\overleftarrow{D}+m_B}{2m_B}\bigg)^{-1}\gamma_\mu\gamma_5A_\mu\bigg(1-\frac{iv\hspace{-0.2cm}\slash
v\cdot
D+m_B}{2m_B}\bigg)^{-1}\frac{iD\hspace{-0.25cm}\slash_\bot}{2m_B}\hat{H}_v\bigg\}
\end{eqnarray}
where
\begin{eqnarray}
&&\hat{H}_v=\hat{H}_v^{(+)}+\hat{H}_v^{(-)}\\
&&\hat{H}_v^{(\pm)}={1\pm v\hspace{-0.2cm}\slash\over2}H^{(\pm)}\\
&&D\hspace{-0.25cm}\slash=D\hspace{-0.2cm}\slash_\parallel+D\hspace{-0.25cm}\slash_\perp\\
&&D\hspace{-0.2cm}\slash_\parallel=v\hspace{-0.2cm}\slash v\cdot
D,~~~~D\hspace{-0.25cm}\slash_\perp=D\hspace{-0.25cm}\slash-v\hspace{-0.2cm}\slash
v\cdot D
\end{eqnarray}
and $H^{(\pm)}$ are the solutions of the Dirac equations
corresponding to the positive and negative energy respectively.
For any operator $O$, the operator $\overleftarrow{O}$ is defined
by $\int \kappa\overleftarrow{O}\varphi\equiv-\int \kappa
O\varphi$.

Defining\cite{HQEFT1} \cite{HQEFT2}
\begin{eqnarray}
A_{\mu\parallel}=v_\mu v\cdot A,~~~~A_{\mu\perp}=A_\mu-v_\mu
v\cdot A
\end{eqnarray}
The Lagrangian can be written in terms of the baryon baryon
($++$),baryon anti-baryon ($+ -$), anti-baryon baryon ($-+$) and
anti-baryon anti-baryon($--$) explicitly as,
\begin{eqnarray}
{\cal L}_v&=&{\cal L}_v^{(++)}+{\cal L}_v^{(--)}+{\cal
L}_v^{(+-)}+{\cal L}_v^{(-+)}\nonumber\\
&&+{\cal L}_{A,v}^{(++)}+{\cal L}_{A,v}^{(--)}+{\cal
L}_{A,v}^{(+-)}+{\cal L}_{A,v}^{(-+)}
\end{eqnarray}
where
\begin{eqnarray}
{\cal L}_v^{(\pm\pm)}&=&{\rm
Tr}[\bar{\hat{H}}_v(i\hat{{\cal D}\hspace{-0.3cm}\slash}_v-m_B)\hat{H}_v]\\
{\cal L}_v^{(\pm\mp)}&=&{1\over2m_B}{\rm
Tr}[\bar{\hat{H}}_v(-i\overleftarrow{D\hspace{-0.25cm}\slash}_\perp)(1-\frac{-iv\hspace{-0.2cm}\slash
v\cdot
\overleftarrow{D}+m_B}{2m_B})^{-1}(i\hat{{\cal D}\hspace{-0.3cm}\slash}_v-m_B)\hat{H}_v]\nonumber\\
{\cal L}_{A,v}^{(\pm\pm)}&=&2D{\rm
Tr}\bigg\{\bar{\hat{H}}_v\gamma_\mu\gamma_5A_{\mu\perp}\hat{H}_v+\bar{\hat{H}}_v\gamma_\mu\gamma_5A_{\mu\parallel}\bigg(1-\frac{iv\hspace{-0.2cm}\slash
v\cdot
D+m_B}{2m_B}\bigg)^{-1}\frac{iD\hspace{-0.25cm}\slash_\bot}{2m_B}\hat{H}_v\nonumber\\
&&+\bar{\hat{H}}_v\frac{-i\overleftarrow{D\hspace{-0.25cm}\slash_\bot}}{2m_B}\bigg(1-\frac{-iv\hspace{-0.2cm}\slash
v\cdot
\overleftarrow{D}+m_B}{2m_B}\bigg)^{-1}\gamma_\mu\gamma_5A_{\mu\parallel}\hat{H}_v\nonumber\\
&&+\bar{\hat{H}}_v\frac{-i\overleftarrow{D\hspace{-0.25cm}\slash_\bot}}{2m_B}\bigg(1-\frac{-iv\hspace{-0.2cm}\slash
v\cdot
\overleftarrow{D}+m_B}{2m_B}\bigg)^{-1}\gamma_\mu\gamma_5A_{\mu\perp}\bigg(1-\frac{iv\hspace{-0.2cm}\slash
v\cdot
D+m_B}{2m_B}\bigg)^{-1}\frac{iD\hspace{-0.25cm}\slash_\bot}{2m_B}\hat{H}_v\bigg\}\nonumber\\
{\cal L}_{A,v}^{(\pm\mp)}&=&2D{\rm
Tr}\bigg\{\bar{\hat{H}}_v\gamma_\mu\gamma_5A_{\mu\parallel}\hat{H}_v+\bar{\hat{H}}_v\gamma_\mu\gamma_5A_{\mu\perp}\bigg(1-\frac{iv\hspace{-0.2cm}\slash
v\cdot
D+m_B}{2m_B}\bigg)^{-1}\frac{iD\hspace{-0.25cm}\slash_\bot}{2m_B}\hat{H}_v\nonumber\\
&&+\bar{\hat{H}}_v\frac{-i\overleftarrow{D\hspace{-0.25cm}\slash_\bot}}{2m_B}\bigg(1-\frac{-iv\hspace{-0.2cm}\slash
v\cdot
\overleftarrow{D}+m_B}{2m_B}\bigg)^{-1}\gamma_\mu\gamma_5A_{\mu\perp}\hat{H}_v\nonumber\\
&&+\bar{\hat{H}}_v\frac{-i\overleftarrow{D\hspace{-0.25cm}\slash_\bot}}{2m_B}\bigg(1-\frac{-iv\hspace{-0.2cm}\slash
v\cdot
\overleftarrow{D}+m_B}{2m_B}\bigg)^{-1}\gamma_\mu\gamma_5A_{\mu\parallel}\bigg(1-\frac{iv\hspace{-0.2cm}\slash
v\cdot
D+m_B}{2m_B}\bigg)^{-1}\frac{iD\hspace{-0.25cm}\slash_\bot}{2m_B}\hat{H}_v\bigg\}
\end{eqnarray}
with
\begin{eqnarray}
i\hat{{\cal D}\hspace{-0.25cm}\slash}_v=iv\hspace{-0.2cm}\slash
v\cdot
D+{1\over2m_B}iD\hspace{-0.25cm}\slash_\perp(1-\frac{iv\hspace{-0.2cm}\slash
v\cdot D+m_B}{2m_B})^{-1}iD\hspace{-0.25cm}\slash_\perp
\end{eqnarray}

To make $1/m_B$ expansion, it is useful to remove the large mass
term in the Lagrangian. Introducing new field variables $H_v$ and
$\hat{H}_v$ with the definition, We can rewrite the above
Lagrangian as\cite{HQEFT1,HQEFT2,HQEFT3}
\begin{eqnarray}
H_v=e^{iv\hspace{-0.15cm}\slash m_B v\cdot
x}\hat{H}_v,~~~~\bar{H}_v=\bar{\hat{H}}_ve^{-iv\hspace{-0.15cm}\slash
m_B v\cdot x}\label{redef}
\end{eqnarray}
We can rewrite the above Lagrangian as,
\begin{eqnarray}
{\cal L}_v^{(\pm\pm)}&=&{\rm
Tr}[\bar{H_v}i\hat{{\cal D}\hspace{-0.3cm}\slash}_vH_v]\\
{\cal L}_v^{(\pm\mp)}&=&{1\over2m_B}{\rm
Tr}[\bar{H_v}(-i\overleftarrow{D\hspace{-0.25cm}\slash}_\perp)(1-\frac{-iv\hspace{-0.2cm}\slash
v\cdot \overleftarrow{D}}{2m_B})^{-1}e^{-2iv\hspace{-0.15cm}\slash
m_B v\cdot x}(i{\cal
D}\hspace{-0.3cm}\slash_v)H_v]\nonumber\\
{\cal L}_{A,v}^{(\pm\pm)}&=&2D{\rm
Tr}\bigg\{\bar{H}_vA\hspace{-0.2cm}\slash_{\perp}\gamma_5H_v\nonumber\\
&&+\bar{H}_vA\hspace{-0.2cm}\slash_{\parallel}\gamma_5\bigg(1-\frac{iv\hspace{-0.2cm}\slash
v\cdot
D}{2m_B}\bigg)^{-1}\frac{iD\hspace{-0.25cm}\slash_\bot}{2m_B}H_v+\bar{H}_v\frac{-i\overleftarrow{D\hspace{-0.25cm}\slash_\bot}}{2m_B}\bigg(1-\frac{-iv\hspace{-0.2cm}\slash
v\cdot
\overleftarrow{D}}{2m_B}\bigg)^{-1}A\hspace{-0.2cm}\slash_{\parallel}\gamma_5H_v\nonumber\\
&&+\bar{H}_v\frac{-i\overleftarrow{D\hspace{-0.25cm}\slash_\bot}}{2m_B}\bigg(1-\frac{-iv\hspace{-0.2cm}\slash
v\cdot
\overleftarrow{D}}{2m_B}\bigg)^{-1}A\hspace{-0.2cm}\slash_{\perp}\gamma_5\bigg(1-\frac{iv\hspace{-0.2cm}\slash
v\cdot
D}{2m_B}\bigg)^{-1}\frac{iD\hspace{-0.25cm}\slash_\bot}{2m_B}H_v\bigg\}\nonumber\\
{\cal L}_{A,v}^{(\pm\mp)}&=&2D{\rm
Tr}\bigg\{\bar{H}_vA\hspace{-0.2cm}\slash_{\parallel}\gamma_5e^{-2iv\hspace{-0.15cm}\slash m_Bv\cdot x}H_v\nonumber\\
&&+\bar{H}_vA\hspace{-0.2cm}\slash_{\perp}\gamma_5\bigg(1-\frac{iv\hspace{-0.2cm}\slash
v\cdot D}{2m_B}\bigg)^{-1}e^{2iv\hspace{-0.15cm}\slash m_Bv\cdot
x}\frac{iD\hspace{-0.25cm}\slash_\perp}{2m_B}H_v+\bar{H}_v\frac{-i\overleftarrow{D\hspace{-0.25cm}\slash_\perp}}{2m_B}\bigg(1-\frac{-iv\hspace{-0.2cm}\slash
v\cdot
\overleftarrow{D}}{2m_B}\bigg)^{-1}\nonumber\\
&&\times A\hspace{-0.2cm}\slash_{\perp}\gamma_5e^{-2iv\hspace{-0.15cm}\slash m_Bv\cdot x}H_v\nonumber\\
&&+\bar{H}_v\frac{-i\overleftarrow{D\hspace{-0.25cm}\slash_\bot}}{2m_B}\bigg(1-\frac{-iv\hspace{-0.2cm}\slash
v\cdot
\overleftarrow{D}}{2m_B}\bigg)^{-1}A\hspace{-0.2cm}\slash_{\parallel}\gamma_5\bigg(1-\frac{iv\hspace{-0.2cm}\slash
v\cdot
D}{2m_B}\bigg)^{-1}\frac{iD\hspace{-0.25cm}\slash_\perp}{2m_B}e^{-2iv\hspace{-0.15cm}\slash
m_Bv\cdot x}H_v\bigg\}
\end{eqnarray}
with
\begin{eqnarray}
i\hat{{\cal D}\hspace{-0.25cm}\slash}_v=iv\hspace{-0.2cm}\slash
v\cdot
D+{1\over2m_B}iD\hspace{-0.25cm}\slash_\perp(1-\frac{iv\hspace{-0.2cm}\slash
v\cdot D}{2m_B})^{-1}iD\hspace{-0.25cm}\slash_\perp
\end{eqnarray}
The factor $e^{\pm 2iv\hspace{-0.15cm}\slash m_Bv\cdot x}$ arises
from the opposite momentum shift for the effective heavy baryon
and anti-baryon fields. Introducing  the electromagnetic field
\begin{eqnarray}
&&D_\mu H_v(x)=\partial_\mu H_v(x)-iV_\mu H_v(x)\\
&&V_\mu(x)={i\over2}[\xi(x)(\partial_\mu-ieQ{\cal
A}_\mu)\xi^\dag(x)+\xi^\dag(x)(\partial_\mu-ieQ{\cal
A}_\mu)\xi(x)]\nonumber\\
&&A_\mu(x)={i\over2}[\xi(x)(\partial_\mu-ieQ{\cal
A}_\mu)\xi^\dag(x)-\xi^\dag(x)(\partial_\mu-ieQ{\cal
A}_\mu)\xi(x)]
\end{eqnarray}
where $Q=(1+\tau_3)/2$ and $\tau_3$ is the third component of the
isospin. In the Coulomb gauge
\begin{eqnarray}
{\cal A}_0=0,~~~~ v\cdot{\cal A}=0
\end{eqnarray}
it should be noticed that the polarization direction introduced in
this paper is different from the convention used
in\cite{HBreview1}.

\subsection{The Leading Order
Lagrangian (${\cal L}_{\pi N}^{(1)}$)}

The lowest order Lagrangian (${\cal L}_{\pi N}^{(1)}$)(where the
superscript (1) denotes the low energy dimension (number of
derivatives and/or quark mass terms))can be decomposed into
\begin{eqnarray}
{\cal L}_v^{(\pm\pm)}&=&{\rm Tr}[\bar{H_v}(iv\hspace{-0.2cm}\slash
v\cdot D)H_v]\nonumber\\
{\cal L}_{A,v}^{(\pm\pm)}&=&2D{\rm
Tr}\{\bar{H}_vA\hspace{-0.2cm}\slash_{\perp}\gamma_5H_v\}\nonumber\\
{\cal L}_{A,v}^{(\pm\mp)}&=&2D{\rm
Tr}\{\bar{H}_vA\hspace{-0.2cm}\slash_{\parallel}\gamma_5e^{-2iv\hspace{-0.2cm}\slash
m_Bv\cdot x}H_v\}
\end{eqnarray}
with $ S_v^\mu$ the covariant spin operator\cite{HBreview1}
\begin{eqnarray}
S_v^\mu={i\over2}\gamma_5\sigma^{\mu\nu}v_\nu=-{1\over2}\gamma_5(\gamma^\mu
v\hspace{-0.2cm}\slash-v^\mu),~~~~S_v^{\mu\dag}=\gamma_0S^\mu_v\gamma_0
\end{eqnarray}
with
\begin{eqnarray}
\sigma^{\mu\nu}={i\over2}[\gamma^\mu,\gamma^\nu]
\end{eqnarray}
The Lagrangian can be expressed as
\begin{eqnarray}
{\cal L}_v^{(\pm\pm)}&=&{\rm Tr}[\bar{H_v}^{(+)}(iv\cdot
D)H_v^{(+)}]-{\rm
Tr}[\bar{H_v}^{(-)}(iv\cdot D)H_v^{(-)}]\label{LO1}\\
{\cal L}_{A,v}^{(\pm\pm)}&=&4D{\rm Tr}\{\bar{H}_v^{(+)}S_v\cdot
A_{\perp}H_v^{(+)}\}-4D{\rm
Tr}\{\bar{H}_v^{(-)}S_v\cdot A_{\perp}H_v^{(-)}\}\label{LO2}\\
{\cal L}_{A,v}^{(\pm\mp)}&=&2D{\rm Tr}\{\bar{H}_v^{(+)}\gamma_5
v\cdot A_{\parallel}e^{2im_Bv\cdot x}H_v^{(-)}\}-2D{\rm
Tr}\{\bar{H}_v^{(-)}\gamma_5v\cdot A_{\parallel}e^{-2im_Bv\cdot
x}H_v^{(+)}\}\label{LO3}
\end{eqnarray}
From the above Lagrangian, we can extract the Feynman rules
(Appendix A) the relations between the constants in our convention
and those given in\cite{HBreview1} are
\begin{eqnarray}
&&f_\pi=2F\\
&&D=-{1\over2}g_A
\end{eqnarray}

\subsection{The Next to Leading Order Lagrangian (${\cal L}_{\pi
N}^{(2)}$)}

In this section we will write down the effective (${\cal L}_{\pi
N}^{(2)}$) Lagrangian.  The terms which stem from the $1/m_B$
expansion of the relativistic $\pi N$ Lagrangian are
\begin{eqnarray}
{\cal L}_{v\pi N{1/m_B}}^{(\pm\pm)}&=&-{1\over2m_B}{\rm
Tr}[\bar{H}_vD\hspace{-0.25cm}\slash_\perp
D\hspace{-0.25cm}\slash_\perp)H_v]\nonumber\\
{\cal L}_{v\pi N{1/m_B}}^{(\pm\mp)}&=&{1\over2m_B}{\rm
Tr}[\bar{H_v}(-i\overleftarrow{D\hspace{-0.25cm}\slash}_\perp)e^{-2iv\hspace{-0.15cm}\slash
m_B v\cdot x}(i{\cal
D}\hspace{-0.3cm}\slash_v)H_v]\nonumber\\
{\cal L}_{A,v,1/{m_B}}^{(\pm\pm)}&=&{2D\over2m_B}{\rm
Tr}\bigg\{\bar{H}_vA\hspace{-0.2cm}\slash_{\parallel}\gamma_5iD\hspace{-0.25cm}\slash_\bot
H_v+\bar{H}_v(-i\overleftarrow{D\hspace{-0.25cm}\slash_\bot})A\hspace{-0.2cm}\slash_{\parallel}\gamma_5H_v\bigg\}\nonumber\\
{\cal L}_{A,v,1/{m_B}}^{(\pm\mp)}&=&{2D\over2m_B}{\rm
Tr}\bigg\{\bar{H}_vA\hspace{-00.2cm}\slash_{\perp}\gamma_5e^{2iv\hspace{-0.2cm}\slash
m_Bv\cdot x}iD\hspace{-0.25cm}\slash_\bot
H_v+\bar{H}_v(-i\overleftarrow{D\hspace{-0.25cm}\slash_\bot})A\hspace{-0.2cm}\slash_{\perp}\gamma_5e^{-2iv\hspace{-0.2cm}\slash
m_Bv\cdot x}H_v\bigg\}
\end{eqnarray}
The other terms involving the low energy constants (LECs) come
from the most general relativistic Lagrangian at $O(p^2)$ after
translation into the heavy mass formalism. There are constants
which appear in the field $\chi$, and is related to explicit
chiral symmetry breaking\cite{HBreview1} These are,
\begin{eqnarray}
{\cal L}_{A\pi N}^{(2)}&=&c_1^\prime\bar{H}H{\rm
Tr}[\chi^+]+c_2^\prime{\rm
Tr}[\bar{H}\chi^+H]+c_3^\prime\bar{H}\sigma^{\mu\nu}H{\rm
Tr}[f_{\mu\nu}^+]+c_4^\prime{\rm
Tr}[\bar{H}\sigma^{\mu\nu}f_{\mu\nu}^+H]\nonumber\\
&&+c^\prime_5{\rm Tr}[\bar{H}\{\gamma_\mu,\gamma_\nu\}A_\mu A_\nu
H]+c^\prime_6{\rm Tr}[\bar{H}[\gamma_\mu,\gamma_\nu]A_\mu A_\nu H]
\end{eqnarray}
where $c_i^\prime$ are the LECs, $\chi^+=\xi^\dag\chi
\xi^\dag+\xi\chi^\dag \xi$, $\chi=2B{\cal M}$ ($\cal M$) is quark
mass matrix and $f^+_{\mu\nu}=e(\partial_\mu{\cal
A}_\nu-\partial_\nu{\cal A}_\mu)(\xi Q\xi^\dag+\xi^\dag Q\xi)$
where $F_{\mu\nu}=(\partial_\mu{\cal A}_\nu-\partial_\nu{\cal
A}_\mu)$ is the canonical photon field strength tensor.
Integrating  out the small component of the field, we get the
$O(p^2)$ Lagrangian
\begin{eqnarray}
{\cal L}_{A\pi N}^{(2)}&=&c_1^\prime\bar{\hat{H}}_v\hat{H}_v{\rm
Tr}[\chi^+]+c_2^\prime{\rm
Tr}[\bar{\hat{H}}_v\chi^+\hat{H}_v]+c_3^\prime\bar{\hat{H}}_v(\partial\hspace{-0.2cm}\slash{\cal
A}\hspace{-0.2cm}\slash-\partial_\mu {\cal A}_\mu)\hat{H}_v{\rm
Tr}[e(\xi Q\xi^\dag+\xi^\dag Q\xi)]\nonumber\\
&&+c_4^\prime{\rm
Tr}[\bar{\hat{H}}_v(\partial\hspace{-0.2cm}\slash{\cal
A}\hspace{-0.2cm}\slash-\partial_\mu {\cal A}_\mu)e(\xi Q\xi^\dag+\xi^\dag Q\xi)\hat{H}_v]\nonumber\\
&&+c^\prime_5{\rm Tr}[\bar{\hat{H}}_vA\cdot
A\hat{H}_v]+c^\prime_6{\rm
Tr}[\bar{\hat{H}}_vA\hspace{-0.2cm}\slash
A\hspace{-0.2cm}\slash\hat{H}_v]
\end{eqnarray}
again expressing the Lagrangian in terms of baryon and anti-baryon
fields we have,
\begin{eqnarray}
{\cal L}_{A\pi
N}^{(2)(\pm,\pm)}&=&c_1^\prime\bar{\hat{H}}_v\hat{H}_v{\rm
Tr}[\chi^+]+c_2^\prime{\rm
Tr}[\bar{\hat{H}}_v\chi^+\hat{H}_v]+c_3^\prime\bar{\hat{H}}_v(\partial\hspace{-0.2cm}\slash_\perp{\cal
A}\hspace{-0.2cm}\slash_\perp-\partial_\mu {\cal
A}_\mu)\hat{H}_v{\rm
Tr}[e(\xi Q\xi^\dag+\xi^\dag Q\xi)]\nonumber\\
&&+c_4^\prime{\rm
Tr}[\bar{\hat{H}}_v(\partial\hspace{-0.2cm}\slash_\perp{\cal
A}\hspace{-0.2cm}\slash_\perp-\partial_\mu {\cal A}_\mu)e(\xi Q\xi^\dag+\xi^\dag Q\xi)\hat{H}_v]\nonumber\\
&&+c^\prime_5{\rm Tr}[\bar{\hat{H}}_vA\cdot
A\hat{H}_v]+c^\prime_6{\rm
Tr}[\bar{\hat{H}}_v(A\hspace{-0.2cm}\slash_\parallel
A\hspace{-0.2cm}\slash_\parallel+A\hspace{-0.2cm}\slash_\perp
A\hspace{-0.2cm}\slash_\perp)\hat{H}_v]\\
{\cal L}_{A\pi
N}^{(2)(\pm,\mp)}&=&c_3^\prime\bar{\hat{H}}_v(\partial\hspace{-0.2cm}\slash_\parallel{\cal
A}\hspace{-0.2cm}\slash_\perp)\hat{H}_v{\rm Tr}[e(\xi
Q\xi^\dag+\xi^\dag Q\xi)]+c_4^\prime{\rm
Tr}[\bar{\hat{H}}_v(\partial\hspace{-0.2cm}\slash_\parallel{\cal
A}\hspace{-0.2cm}\slash_\perp)e(\xi Q\xi^\dag+\xi^\dag Q\xi)\hat{H}_v]\nonumber\\
&&+c^\prime_6{\rm
Tr}[\bar{\hat{H}}_v(A\hspace{-0.2cm}\slash_\parallel
A\hspace{-0.2cm}\slash_\perp+A\hspace{-0.2cm}\slash_\perp
A\hspace{-0.2cm}\slash_\parallel)\hat{H}_v]
\end{eqnarray}

Following  the definition $(\ref{redef})$, the Lagrangian can be
rewritten as
\begin{eqnarray}
{\cal L}_{A\pi N}^{(2)(\pm,\pm)}&=&c_1^\prime\bar{H}_vH_v{\rm
Tr}[\chi^+]+c_2^\prime{\rm
Tr}[\bar{H}_v\chi^+H_v]+c_3^\prime\bar{H}_v(\partial\hspace{-0.2cm}\slash_\perp{\cal
A}\hspace{-0.2cm}\slash_\perp-\partial_\mu {\cal A}_\mu)H_v{\rm
Tr}[e(\xi Q\xi^\dag+\xi^\dag Q\xi)]\nonumber\\
&&+c_4^\prime{\rm
Tr}[\bar{H}_v(\partial\hspace{-0.2cm}\slash_\perp{\cal
A}\hspace{-0.2cm}\slash_\perp-\partial_\mu {\cal A}_\mu)e(\xi Q\xi^\dag+\xi^\dag Q\xi)H_v]\nonumber\\
&&+c^\prime_5{\rm Tr}[\bar{H}_vA\cdot AH_v]+c^\prime_6{\rm
Tr}[\bar{H}_v(A\hspace{-0.2cm}\slash_\parallel
A\hspace{-0.2cm}\slash_\parallel+A\hspace{-0.2cm}\slash_\perp
A\hspace{-0.2cm}\slash_\perp)H_v]\\
{\cal L}_{A\pi
N}^{(2)(\pm,\mp)}&=&c_3^\prime\bar{H}_v(\partial\hspace{-0.2cm}\slash_\parallel{\cal
A}\hspace{-0.2cm}\slash_\perp)e^{-2iv\hspace{-0.15cm}\slash m_B
v\cdot x}H_v{\rm Tr}[e(\xi Q\xi^\dag+\xi^\dag
Q\xi)]\nonumber\\
&&+c_4^\prime{\rm
Tr}[\bar{H}_v(\partial\hspace{-0.2cm}\slash_\parallel{\cal
A}\hspace{-0.2cm}\slash_\perp)e(\xi Q\xi^\dag+\xi^\dag
Q\xi)e^{-2iv\hspace{-0.15cm}\slash m_B v\cdot
x}H_v]\nonumber\\
&&+c^\prime_6{\rm Tr}[\bar{H}_v(A\hspace{-0.2cm}\slash_\parallel
A\hspace{-0.2cm}\slash_\perp+A\hspace{-0.2cm}\slash_\perp
A\hspace{-0.2cm}\slash_\parallel)e^{-2iv\hspace{-0.15cm}\slash m_B
v\cdot x}H_v]
\end{eqnarray}

The  $O(p^2)$ Lagrangian without baryon-antibaryon mixing is
\begin{eqnarray}
{\cal L}_{v\pi N{1/m_B}}^{(\pm\pm)}&=&-{1\over2m_B}{\rm
Tr}[D\hspace{-0.25cm}\slash_\perp
D\hspace{-0.25cm}\slash_\perp)H_v]\nonumber\\
{\cal L}_{A,v,1/{m_B}}^{(\pm\pm)}&=&{2D\over2m_B}{\rm
Tr}\bigg\{\bar{H}_vA\hspace{-0.2cm}\slash_{\parallel}\gamma_5iD\hspace{-0.25cm}\slash_\bot
H_v+\bar{H}_v(-i\overleftarrow{D\hspace{-0.25cm}\slash_\bot})A\hspace{-0.2cm}\slash_{\parallel}\gamma_5H_v\bigg\}\nonumber\\
{\cal L}_{A\pi N}^{(2)(\pm,\pm)}&=&c_1^\prime\bar{H}_vH_v{\rm
Tr}[\chi^+]+c_2^\prime{\rm
Tr}[\bar{H}_v\chi^+H_v]+c_3^\prime\bar{H}_v(\partial\hspace{-0.2cm}\slash_\perp{\cal
A}\hspace{-0.2cm}\slash_\perp-\partial_\mu {\cal A}_\mu)H_v{\rm
Tr}[e(\xi Q\xi^\dag+\xi^\dag Q\xi)]\nonumber\\
&&+c_4^\prime{\rm
Tr}[\bar{H}_v(\partial\hspace{-0.2cm}\slash_\perp{\cal
A}\hspace{-0.2cm}\slash_\perp-\partial_\mu {\cal A}_\mu)e(\xi Q\xi^\dag+\xi^\dag Q\xi)H_v]\nonumber\\
&&+c^\prime_5{\rm Tr}[\bar{H}_vA\cdot AH_v]+c^\prime_6{\rm
Tr}[\bar{H}_v(A\hspace{-0.2cm}\slash_\parallel
A\hspace{-0.2cm}\slash_\parallel+A\hspace{-0.2cm}\slash_\perp
A\hspace{-0.2cm}\slash_\perp)H_v]
\end{eqnarray}

The $O(p^2)$ Lagrangian with baryon-antibaryon mixing is
\begin{eqnarray}
{\cal L}_{v\pi N{1/m_B}}^{(\pm\mp)}&=&{1\over2m_B}{\rm
Tr}[\bar{H_v}(-i\overleftarrow{D\hspace{-0.25cm}\slash}_\perp)e^{-2iv\hspace{-0.15cm}\slash
m_B v\cdot x}(iv\hspace{-0.2cm}\slash v\cdot D)H_v]\nonumber\\
{\cal L}_{A,v,1/{m_B}}^{(\pm\mp)}&=&{2D\over2m_B}{\rm
Tr}\bigg\{\bar{H}_vA\hspace{-00.2cm}\slash_{\perp}\gamma_5e^{2iv\hspace{-0.2cm}\slash
m_Bv\cdot x}iD\hspace{-0.25cm}\slash_\bot
H_v+\bar{H}_v(-i\overleftarrow{D\hspace{-0.25cm}\slash_\bot})A\hspace{-0.2cm}\slash_{\perp}\gamma_5e^{-2iv\hspace{-0.2cm}\slash
m_Bv\cdot x}H_v\bigg\}\nonumber\\
{\cal L}_{A\pi
N}^{(2)(\pm,\mp)}&=&c_3^\prime\bar{H}_v(\partial\hspace{-0.2cm}\slash_\parallel{\cal
A}\hspace{-0.2cm}\slash_\perp)e^{-2iv\hspace{-0.15cm}\slash m_B
v\cdot x}H_v{\rm Tr}[e(\xi Q\xi^\dag+\xi^\dag
Q\xi)]\nonumber\\
&&+c_4^\prime{\rm
Tr}[\bar{H}_v(\partial\hspace{-0.2cm}\slash_\parallel{\cal
A}\hspace{-0.2cm}\slash_\perp)e(\xi Q\xi^\dag+\xi^\dag
Q\xi)e^{-2iv\hspace{-0.15cm}\slash m_B v\cdot x}H_v]
\end{eqnarray}
In the above equations the Dirac matrices can be expressed as a
combination of the $S_v^\mu$, four velocity $v_\mu$ and
$\gamma_5$. It is to be noted that the relations between the
constants in our convention and
 that of  \cite{HBreview1} are
\begin{eqnarray}
&&c_3^\prime={1\over2}[\frac{i\kappa_s}{4m_B}-\frac{i\kappa_v}{4m_B}]\\
&&c_4^\prime=\frac{i\kappa_v}{4m_B}
\end{eqnarray}
where $\kappa_s$ and $\kappa_v$ are the scalar and vector
anomalous magnetic moments.


\section{Spin Polarisability of the
 Nucleon in the frame work of HBEFT }

In this section, we  use the Lagrangian constructed in previous
section to calculate the polarisabilities of the nucleon.

To calculate the spin-dependent scattering amplitude, we work in
the gauge $A_0=0$ ($v\cdot A = 0 $), or in the language of HBCHPT,
$\epsilon\cdot v=0$, where $v^\mu =(1,0,0,0)$ is the unit vector
which defines the nucleon rest frame. Here, there is no
lowest-order coupling of a photon to a nucleon; the coupling comes
in only at second order. The Feynman vertex consists of two
pieces, one proportional to the charge current and one to the
magnetic moment
\begin{eqnarray}
\frac{ie}{2m_B}\{Q\epsilon\cdot(p_1+p_2)+2(Q+\kappa)[S\cdot
\epsilon, S\cdot Q]\}
\end{eqnarray}

\begin{figure}[htbp]
\begin{center}
\epsfxsize = 14cm \  \epsfbox{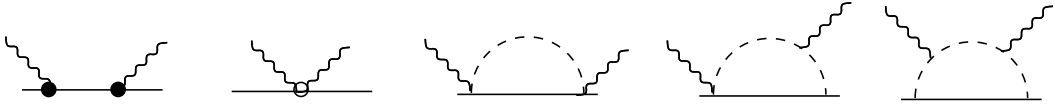}
\end{center}
\caption[Diagrams which contribute to spin-dependent
Compton Scattering in the $\epsilon\cdot v=0$ gauge at LO. The open circles are vertices
from  ${\cal L}_{\pi N}^{(2)}$ and the solid dot is a vertex from  ${\cal L}_{\pi N}^{(3)}$]{%
Diagrams which contribute to spin-dependent Compton Scattering in
the $\epsilon\cdot v=0$ gauge at LO. } \label{fig:LO}
\end{figure}

At leading order\cite{Op22}, the diagrams in Fig.~\ref{fig:LO}
should be considered.  At the leading order LET is satisfied by
the combination of the Born and seagull diagrams. The calculated
polarisabilities agrees with those of ref.\cite{Op22}.

At NLO, the diagrams which contribute to spin dependent forward
Compton scattering are given in Fig.2 and Fig.3 respectively.
Fig.2 are the diagrams which arise from the $O(p^2)$Lagrangian
without baryon anti baryon mixing (eq.2.45).The amplitude for the
diagrams 2a-2h are listed in appendix B. At this order there can
be no seagulls \cite{Op42}.
\begin{figure}[htbp]
\begin{center}
\epsfxsize = 14cm \  \epsfbox{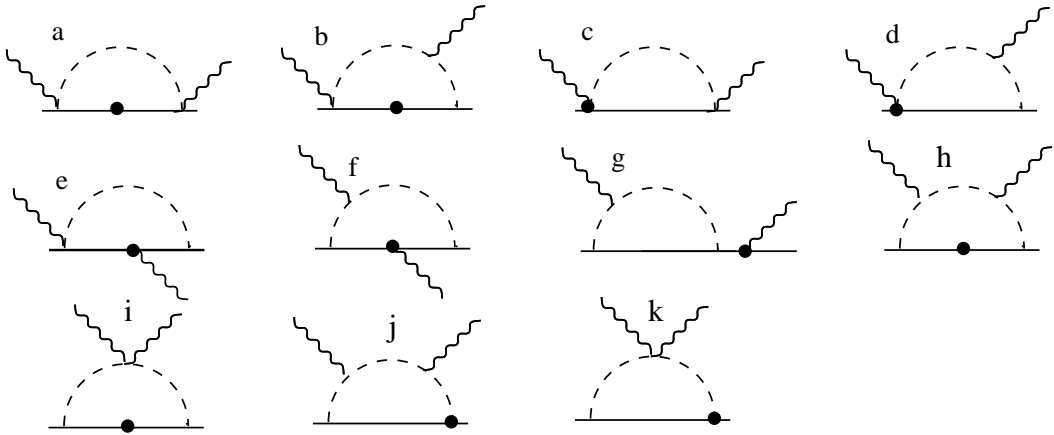}
\end{center}
\caption[Diagrams which contribute to spin-dependent
Compton Scattering in the $\epsilon\cdot v=0$ gauge at NLO. The solid dots
are vertices from ${\cal L}_{\pi N}^{(2)}$
]{%
Diagrams which contribute to spin-dependent Compton Scattering in
the $\epsilon\cdot v=0$ gauge at NLO. These are the diagrams from
the Lagrangian without baryon anti baryon mixing (crossed diagrams
are not shown).} \label{fig:NLO}
\end{figure}
In Fig.~\ref{fig:NLO}, the solid dots are vertices from ${\cal
L}_{\pi N}^{(2)}$. The amplitudes for the loop diagrams are listed
in Appendix B.

Without considering the resonance contributions,
$A_i(\omega,\theta)$ can be divided into the following structure
\begin{eqnarray}
A_i(\omega,\theta)=A_i(\omega,\theta)^{Born}+A_i(\omega,\theta)^{HBChPT}+A_i(\omega,\theta)^{antibaryon}\label{Wdecomp}
\end{eqnarray}
The first two terms in the above expression have been worked out
in  the framework of HBCHPT by number of authors. Our results of
$\gamma_i$ naturally agrees with those of ref.
\cite{Op42}\cite{Op43}:
\begin{eqnarray}
\gamma_1^{HBChPT}&=&\frac{\alpha_{em}g^2_A}{24\pi^2F^2
m^2_\pi}[1-\frac{\pi m_\pi}{8m_B}(8+5\tau_3)]\nonumber\\
\gamma_2^{HBChPT}&=&\frac{\alpha_{em}g^2_A}{48\pi^2F^2
m^2_\pi}[1-\frac{\pi
m_\pi}{4m_B}(8+\kappa_v+3(1+\kappa_s)\tau_3)]\nonumber\\
\gamma_3^{HBChPT}&=&\frac{\alpha_{em}g^2_A}{96\pi^2F^2
m^2_\pi}[1-\frac{\pi
m_\pi}{4m_B}(6+\tau_3)]\nonumber\\
\gamma_4^{HBChPT}&=&\frac{\alpha_{em}g^2_A}{96\pi^2F^2
m^2_\pi}[-1+\frac{\pi
m_\pi}{4m_B}(15+4\kappa_v+4(1+\kappa_s)\tau_3)]\nonumber\\
\gamma_5^{HBChPT}&=&\frac{\alpha_{em}g^2_A}{24\pi^2F^2
m^2_\pi}[-\frac{\pi
m_\pi}{8m_B}(7+3\kappa_v+(1+\kappa_s)\tau_3)]\nonumber\\
\gamma_0^{HBChPT}&=&\frac{\alpha_{em}g^2_A}{24\pi^2F^2
m^2_\pi}[1-\frac{\pi
m_\pi}{8m_B}(15+3\kappa_v+(6+\kappa_s)\tau_3)]\nonumber\\
\gamma_\pi^{HBChPT}&=&\frac{\alpha_{em}g^2_A}{24\pi^2F^2
m^2_\pi}[1-\frac{\pi
m_\pi}{8m_B}(1-3\kappa_v+(4-\kappa_s)\tau_3)]\label{gamma_i}
\end{eqnarray}
where $F$ is the pion decay constant. To get the above relations,
the relations
$\gamma_\theta=\gamma_1-(\gamma_2+2\gamma_4)\cos\theta$ and
$\gamma_5+\gamma_2+2\gamma_4=0$ were used.

At NLO, the diagrams which contribute to the scattering amplitude
from the baryon and anti-baryon vertex are given in Fig.3. These
are the diagrams which arise from the $O(p^2)$ Lagrangian with
baryon-anti baryon mixing (eq.2.46). They have no analog in
HBCHPT.

\begin{picture}(600,60)(0,0)
\Line(35,20)(105,20)\DashCArc(70,20)(20,0,180){2}\Photon(53,30)(35,50){2}{5}\Photon(100,20)(100,50){2}{6}\Text(100,20)[]{$\bullet$}\Text(70,10)[]{$+$}\Text(95,10)[]{$-$}\Text(70,2)[]{$3a$}
\Line(135,20)(205,20)\DashCArc(170,20)(20,0,180){2}\Photon(150,20)(135,50){2}{6}\Photon(170,20)(205,50){2}{7}\Text(170,20)[]{$\bullet$}\Text(160,10)[]{$+$}\Text(180,10)[]{$-$}\Text(170,2)[]{$3b$}
\Line(235,20)(305,20)\DashCArc(270,20)(20,0,180){2}\Photon(253,30)(235,50){2}{5}\Photon(270,20)(305,50){2}{7}\Text(270,20)[]{$\bullet$}\Text(260,10)[]{$+$}\Text(280,10)[]{$-$}\Text(270,2)[]{$3c$}
\Line(335,20)(405,20)\DashCArc(370,20)(20,0,180){2}\Photon(350,20)(335,50){2}{6}\Photon(400,20)(400,50){2}{6}\Text(400,20)[]{$\bullet$}\Text(370,10)[]{$+$}\Text(395,10)[]{$-$}\Text(370,2)[]{$3d$}
\end{picture}
\begin{picture}(600,60)(0,0)
\Line(35,20)(105,20)\DashCArc(70,20)(20,0,180){2}\Photon(40,20)(40,50){2}{6}\Photon(87,30)(100,50){2}{5}\Text(40,20)[]{$\bullet$}\Text(40,10)[]{$-$}\Text(70,10)[]{$+$}\Text(70,2)[]{$3e$}
\Line(135,20)(205,20)\DashCArc(170,20)(20,0,180){2}\Photon(150,20)(135,50){2}{6}\Photon(170,20)(205,50){2}{7}\Text(170,20)[]{$\bullet$}\Text(160,10)[]{$-$}\Text(180,10)[]{$+$}\Text(170,2)[]{$3f$}
\Line(235,20)(305,20)\DashCArc(270,20)(20,0,180){2}\Photon(253,30)(235,50){2}{5}\Photon(270,20)(305,50){2}{7}\Text(270,20)[]{$\bullet$}\Text(260,10)[]{$-$}\Text(280,10)[]{$+$}\Text(270,2)[]{$3g$}
\Line(335,20)(405,20)\DashCArc(370,20)(20,0,180){2}\Photon(350,20)(335,50){2}{6}\Photon(400,20)(400,50){2}{6}\Text(400,20)[]{$\bullet$}\Text(370,10)[]{$-$}\Text(395,10)[]{$+$}\Text(370,2)[]{$3h$}
\end{picture}
\begin{center}{Fig3. The diagrams arising from baryon  antibaryon vertex
which contribute to spin dependent forward Compton scattering at
NLO. The solid dots are vertices from  ${\cal L}_{\pi N}^{(2)}$
(crossed diagrams are not shown)}
\end{center}
In Fig.3 $\pm$ denotes baryon and anti-baryon vertex respectively.

The contribution to forward spin polarisability from the diagrams
of Fig.3 is,
\begin{eqnarray}
T^{O(p^4),anti-baryon}&=&i\frac{\omega^3e^2D^2}{3\pi m_Bm_\pi
f^2_\pi}\{2\kappa_s+1\}\vec{\sigma}\cdot(\vec{\epsilon}^\prime\times\vec{\epsilon})(t_3)+O(\omega^5)\\\label{Tanti-B}
&&+\mbox{spin independent terms}.\nonumber
\end{eqnarray}

At this stage we would wish to make some comments. The LET is
satisfied in our theory as the diagrams of Fig.2 which are same as
that of HBCHPT satisfy the LET. There is no contribution from the
diagrams of Fig.3 to the lowest order in $\omega$ and hence the
LET is intact which is a non-trivial check to our theory. But, the
diagrams of Fig.3 contribute to forward spin polarisability (see
Eq.(3.4)). It should be noted that higher order terms in $\omega$
are model dependent quantities. Also,  the baryon anti-baryon
vertex gives contribution only to $A_3(\omega,\theta)$ and
further, the relation $\gamma_5+\gamma_2+2\gamma_4=0$ is still
exact, which is an another good check to consistency  of our
theory. The explicit contributions from the individual diagrams of
Fig.3 are given in appendix B.

We have
\begin{eqnarray}
A_3^{O(p^4),anti-baryon}(\omega,\theta)&=&\frac{\omega^3e^2D^2}{3\pi
m_Bm_\pi
f^2_\pi}\{2\kappa_s+1\}(t_3)+O(\omega^5)\nonumber\\
&=&\frac{\omega^3e^2g_A^2}{48\pi m_Bm_\pi
F^2}\{2\kappa_s+1\}(t_3)+O(\omega^5)
\end{eqnarray}
we  get
\begin{eqnarray}
\gamma_1^{O(p^4),anti-baryon}(\omega,\theta)&=&\frac{\alpha_{em}g_A^2}{48\pi
m_Bm_\pi F^2}\{2\kappa_s+1\}(t_3)\label{Ngamma_1}
\end{eqnarray}

Using eqs. $(\ref{gamma_i})$ to $(\ref{Ngamma_1})$, we get,
\begin{eqnarray}
\gamma_1&=&\frac{\alpha_{em}g^2_A}{24\pi^2F^2
m^2_\pi}\{1-\frac{\pi
m_\pi}{8m_B}[8+(1-8\kappa_s)\tau_3]\}\nonumber\\
\gamma_0&=&\frac{\alpha_{em}g^2_A}{24\pi^2F^2
m^2_\pi}\{1-\frac{\pi
m_\pi}{8m_B}[15+3\kappa_v+(2+7\kappa_s)\tau_3]\}\nonumber\\
\gamma_\pi&=&\frac{\alpha_{em}g^2_A}{24\pi^2F^2
m^2_\pi}\{1-\frac{\pi m_\pi}{8m_B}[1-3\kappa_v-9\kappa_s\tau_3]\}
\end{eqnarray}

Using  the physical values of the parameter $\kappa_v=3.71$,
$\kappa_s=-0.12$, we  get
\begin{eqnarray}
\gamma_1&=&4.5-(2.0+0.17\tau_3)\nonumber\\
\gamma_0&=&4.5-(6.6+0.29\tau_3)\nonumber\\
\gamma_\pi&=&4.5-(-2.6+0.27\tau_3)
\end{eqnarray}

Below,  we  present our numerical results in Table.1.
\begin{center}
\begin{tabular}{|c|c|c|c|c|}
  \hline
   & $\gamma_0^p$  & $\gamma_\pi^p$ & $\gamma_\pi^n$ \\
  \hline
  PRESENT & $-2.4$  & $-35.8$ & $-34.3$ \\
   \hline
  HBCHPT\cite{Op42}\cite{Op43} & $-3.9$  & $-36.6$ & $-34.4$ \\
   \hline
  Expt & $-0.86\pm0.13$\cite{expe1},  & $-38.7\pm1.8$\cite{expe1} & $-27.1\pm3.6$\cite{expe2} \\
  \hline
\end{tabular}
\\
Table.1. Numerical results of the real Compton scattering in
comparsion with the results of HBCHPT and the extracted results
from the expt (in unit $10^{-4}fm^4$).
\end{center}

From the above table, the following conclusions can be drawn: The
amplitude of the forward spin polarisability $\gamma_0$ of proton
is smaller than the corresponding HBCHPT result, and closer to the
experimental data.  For the backward scattering $\gamma_\pi$, both
calculations lead to similar results which are consistent with the
experimental data whin the errors\cite{expe2}. Since spin
polarisabilities of neutron are sensitive to the
resonance\cite{resonance,HBreview1} and there is no direct
experimental data yet, we shall not consider them in this note. To
further compare our theoretical framework with HBCHPT, it is
required to extend the calculations to the virtual Compton
scattering which is under way\cite{virtual}.

\section{Summary and Conclusions}

In this paper, we have constructed  a complete HBEFT (${\cal
L}_{\pi N}^{(1)}$ and ${\cal L}_{\pi N}^{(2)}$) by considering the
anti-baryon contributions through the idea of HQEFT with photon as
an external degree of freedom. The pure baryon sector of the
derived Lagrangian  is same as the usual HBCHPT Lagrangian.  In
addition the  Lagrangian of HBEFT also consists of the antibaryon
contribution explicitly.

The calculated  spin polarisability at order $O(p^3)$, agree with
the corresponding results of the HBCHPT since from the chiral
power counting, the vertex of $O(p^3)$ loop diagrams  stems from
the insertions of ${\cal L}_{\pi N}^{(1)}$ Lagrangian. This agrees
with the conclusion given in\cite{HQEFT2,HQEFT3} that the HQEFT is
same as the usual HQET at the leading order. But, at the NLO
 the loop diagrams involve one insertion from ${\cal L}_{\pi N}^{(2)}$.
  The calculation shows that  in the frame work of HBEFT
  the result
of $\gamma_1$, is different from the corresponding calculation of
the HBCHPT.  The anti-baryon terms reduces the magnitude of the
spin polarisability of proton at NLO in comparison to the results
of the HBCHPT.

At NLO, the leading order terms in the $\omega$ expansion satisfy
the low energy theorems and the relation
$\gamma_5+\gamma_2+2\gamma_4=0$. Both of the above results  give
very good checks to our theory. Hence, the HBEFT, obtained by
using the idea of the HQEFT is more complete than that of HBCHPT.
If we ignore the diagrams of (Fig.3) (arising from baryon
anti-baryon vertex), our results agree with the results of HBCHPT
\cite{Op42},\cite{Op43}. At NLO, the diagrams of Fig.3 do not
contribute at the leading order in $\omega$ and hence low energy
theorem are satisfied by the leading order terms in $\omega$ of
Fig.2. But, the diagrams of Fig.3 do contribute to spin
polarisabilities.

\acknowledgments

The authors YLMa and YLWu would like to thank Dr. J.P. Chen of
Jefferson Lab for the valuable discussions. This work was
supported in part by the key projects of Chinese Academy of
Sciences(CAS), the National Science Foundation of China (NSFC).The
other author(KBV) would like to thank for providing the excellent
working facilities at ITP, CAS. This work is supported in part by
the TWAS-UNESCO program of ICTP, Italy and the Chinese Academy of
Sciences.

\appendix

\section{Feynman Rules}

 The Lagrangian in the Sec.II denotes the baryon and
anti-baryon explicitly. Below we write down the Feynman rules in
terms of the  baryon fields.  To get the baryon or anti-baryon
contributions, one just need to multiply the velocity projection
operator  $(1+v\hspace{-0.25cm}\slash)/2$ or
$(1-v\hspace{-0.25cm}\slash)/2$ on both sides. For completness we
list the Feynman rules  needed to calculate the tree and loop
diagrams.

We use the following notations:

p Momentum of a nucleon in heavy mass formulation.

q Momentum of an external pion.

k Momentum of an external photon.

$\epsilon$ Photon polarisation vector.

$\hat{x}$ Co-ordinate operator.

Pion isospin indices are a and b. $v_\mu$ is the nucleon
four-velocity and $s_\mu$ is the covariant spin operator.
Parameters Q,D,$f_\pi$... are meant to be taken in the chiral
limit.

(1). The vertices from   ${\cal L}_{\pi N}^{(1)}$ Lagrangian
eqs.$(\ref{LO1})-(\ref{LO3})$.

\begin{picture}(700,150)(0,0)
\Text(40,120)[]{\footnotesize{baryon
propagator}:}\ArrowLine(100,120)(200,120) \Text(300,120)[]
{$\frac{iv\hspace{-0.15cm}\slash}{v\cdot
p}\delta_{ij}$}\Text(150,130)[]{$p$}
\Text(30,80)[]{\footnotesize{1 pion ($q$
out)}:}\DashLine(150,70)(150,100){5}
\Text(145,85)[]{$\uparrow$}\Text(140,85)[]{$q$}\Text(160,100)[]{$a$}
\Line(100,70)(200,70)\Text(300,80)[]{$\frac{-4Dv\hspace{-0.15cm}\slash}{f_\pi}[S_v\cdot
q](t_a)_{ij}$}
\Text(300,60)[]{$-[{1+v\hspace{-0.15cm}\slash\over2}e^{2im_Bv\cdot\hat{x}}-{1-v\hspace{-0.15cm}\slash\over2}e^{-2im_Bv\cdot\hat{x}}]\frac{2D}{f_\pi}\gamma_5v\cdot
q(t_a)_{ij}$}
\Text(35,30)[]{\footnotesize{1 pion, 1
photon}:}\DashLine(150,20)(120,50){5}\Text(130,50)[]{$a$}\Photon(170,50)(150,20){3}{6}
\Line(100,20)(200,20)\Text(300,20)[]{$\frac{4ieDv\hspace{-0.15cm}\slash}{f_\pi}S_v\cdot
\epsilon\epsilon^{a3b}(t_b)_{ij}$}
\end{picture}

(2). The vertices from the ${\cal L}_{\pi N}^{(2)}$ Lagrangian
are.

\begin{picture}(700,150)(0,0)
\Text(40,140)[]{\footnotesize{Baryon propagator}:}
\Line(70,120)(170,120) \Text(140,120)[]{$\rightarrow$}
\Text(120,120)[]{$\bullet$}\Text(315,130)[]
{${-i\over2m_B}[p^2-(v\cdot
p)^2]\delta_{ij}-4iB[c_1^\prime(m_u+m_d)\delta_{ij}+c_2^\prime{\cal
M}_{ij}]$}\Text(140,130)[]{$p$} \Text(310,110)[]{$+{-i\over
m_B}[v\cdot pS_v\cdot
p\gamma_5\delta_{ij}][{1-v\hspace{-0.15cm}\slash\over2}e^{2im_Bv\cdot\hat{x}}+{1+v\hspace{-0.15cm}\slash\over2}e^{-2im_Bv\cdot\hat{x}}]$}
\Text(40,80)[]{\footnotesize{1 Photon ($k$ out)}:}
\Line(70,60)(170,60)
\Text(100,60)[]{$\rightarrow$}\Text(100,70)[]{$p^\prime$}\Text(140,60)[]{$\rightarrow$}\Text(140,70)[]{$p$}
\Photon(120,60)(120,90){3}{6}
\Text(115,75)[]{$\uparrow$}\Text(110,75)[]{$k$}
\Text(120,60)[]{$\bullet$}\Text(330,80)[] {${-
ie\over2m_B}[(p+p^\prime)\cdot\epsilon-2[S_v\cdot
k,S_v\cdot\epsilon]]Q_{ij}$} \Text(330,65)[]{$+4e[S_v\cdot
k,S_v\cdot\epsilon](c_3^\prime+c^\prime_4Q)_{ij}$} \Text(330,50)[]
{$+ieS_v\cdot\epsilon\gamma_5\{4iv\cdot
k[c^\prime_3\delta_{ij}+c^\prime_4Q_{ij}]-{1\over m_B}v\cdot p
Q_{ij}\}$}\Text(330,35)[]
{$\times[{1-v\hspace{-0.15cm}\slash\over2}e^{2im_Bv\cdot\hat{x}}+{1+v\hspace{-0.15cm}\slash\over2}e^{-2im_Bv\cdot\hat{x}}]$}
\end{picture}

It should be noticed that coordinate operator  $\hat{x}$ only acts
on the baryon and antibaryon moments.

\section{Full amplitude}

\subsection{The full amplitude for the diagrams 2a-2h(including the cross ones).
These are the same diagrams which arise at NLO in HBCHPT.
\cite{Op42})}

The notation $t_i$ is used for the tensor structures which
multiply the amplitudes $A_i$\cite{Op42}.
\begin{eqnarray}
T_a&=&{g^2 e^2\over 4m_B f_\pi^2}
\Bigl[\Bigl(m_\pi^2-\omega^2(1+\cos\theta)\Bigr) {\partial
J_0(\omega,m_\pi^2)\over\partial \omega}
-2\omega J_0(\omega,m_\pi^2)\Bigl]t_3 - (\omega\to-\omega)\nonumber\\
T_b&=&-{g^2 e^2\over 2m_B f_\pi^2}
\Bigl[{2m_\pi^2\over\omega}\Bigl(J_2'(\omega,m_\pi^2)-J_2'(0,m_\pi^2)\Bigr)t_3+
(1+\cos\theta){\partial J_2(\omega,m_\pi^2)\over\partial\omega}t_3\nonumber\\
&&-\omega J_2'(\omega,m_\pi^2)t_5+ \omega
\Bigl(t_5-2(1-\cos\theta)t_3\Bigr)\int_0^1\!dx J_2'(x
\omega,m_\pi^2)\Bigr]- (\omega\to-\omega)\nonumber\\
T_c&=&-{g^2 e^2\over 2m_B f_\pi^2}\tau_3\omega
J_0(\omega,m_\pi^2)t_3
- (\omega\to-\omega)\nonumber\\
T_d&=&{g^2 e^2\over m_B f_\pi^2}\tau_3\omega \int_0^1\!dx
J_2'(x\omega,m_\pi^2)t_3
- (\omega\to-\omega)\nonumber\\
T_e&=&{g^2 e^2\over 2m_B f_\pi^2}(1-\tau_3){1\over\omega} \Big(
J_2(\omega,m_\pi^2)-J_2(0,m_\pi^2)\Big)t_3
- (\omega\to-\omega)\nonumber\\
T_f&=&{g^2 e^2\over4 m_B f_\pi^2}\omega
\Bigl(2(\mu_v-\mu_s\tau_3)(t_3\cos\theta-t_4)+(1-\tau_3)t_6\Bigr)\nonumber\\
&&\times\int_0^1\! dx (1-2x)J_2'(x\omega,m_\pi^2)
- (\omega\to-\omega)\nonumber\\
T_g&=&-{g^2 e^2\over 4m_B f_\pi^2}\omega
\Bigl(2(\mu_v+\mu_s\tau_3)(t_3\cos\theta+t_4-t_5)+(1+\tau_3)t_6\Bigr)\nonumber\\
&&\times\int_0^1 \!dx J_2'(x\omega,m_\pi^2)
- (\omega\to-\omega)\nonumber\\
T_h&=&-{g^2 e^2\over m_B f_\pi^2}\omega^2\int_0^1 \!dy
\int_0^{1-x}\!dx\Bigl[ \Big((7x-1)(t_6-t_5)+7(1-x-y)t_4\Bigr)
{\partial J_6''(\tilde\omega,m_\pi^2-xyt)\over\partial\tilde\omega}\nonumber\\
&&+\Bigl(2V(x,y,\theta)(xt_6-xt_5+(1-x-y)t_4)\nonumber\\
&&\;\;\;\;\;\;-(1-x-y)(9xy-x-y)t_7\Big) \omega^2{\partial
J_2''(\tilde\omega,m_\pi^2-xyt)\over\partial\tilde\omega}\nonumber\\
&&-xy(1-x-y)\omega^4V(x,y,\theta)t_7 {\partial
J_0''(\tilde\omega,m_\pi^2-xyt)\over\partial\tilde\omega}\Big] -
(\omega\to-\omega)
\end{eqnarray}
where $\tilde\omega=(1-x-y)\omega$,
\begin{eqnarray}
J_6(\omega,m_\pi^2)={1\over
d+1}\Big((m_\pi^2-\omega^2)J_2(\omega,m_\pi^2) -{\omega
m_\pi^2\over d}\Delta_\pi\Bigr),
\end{eqnarray}
The diagrams 2i-2k does not contribute to spin dependent Compton
scattering and hence are not listed.

The $J_0(\omega,m_\pi^2)$, $J_2(\omega,m_\pi^2)$ and $\Delta_\pi$
are defined in \cite{HBreview1}, prime denotes differentiation
with respect to $m_\pi^2$, and
\begin{eqnarray}
V(x,y,\theta)=(2xy-x-y+1)\cos\theta-x(1-x)-y(1-y).
\end{eqnarray}

\subsection{ Contributions to the Amplitudes from Fig.3 (including the crossed ones)}

\begin{eqnarray}
T_3a&=&-2i\omega C\int_0^1dx[S_v\cdot \epsilon^\prime,S_v\cdot \epsilon]\{8i[c_3^\prime(t_3)_{ti}+c_4^\prime Q_{ti}]-{1\over m_B}Q_{ti}\}\nonumber\\
&&\;\;\;\;\;\;\;\;\;\;\;\;\;\;\;\;\;\;\;\;\;\;\;\;\;\;\;\;\;\;\;\;\;\;\;\;\;\;\;\;\;\times[J_2^\prime(x\omega,m^2_\pi)+J_2^\prime(-x\omega,m^2_\pi)]\nonumber\\
T_3b&=&\frac{i\omega C}{2}[S_v\cdot\epsilon^\prime,
S_v\cdot\epsilon]\{8[2ic_3^\prime(t_3)_{ti}-ic_4^\prime
(1-t_3)_{ti}]+{1\over
m_B}(1-t_3)_{ti}\}\nonumber\\
&&\;\;\;\;\;\;\;\;\;\;\;\;\;\;\;\;\;\;\;\;\;\;\;\;\;\;\;\;\times[J_0(\omega,m^2_\pi)+J_0(-\omega,m^2_\pi)]\nonumber\\
T_3c&=&-i\omega C\int_0^1dx[S_v\cdot\epsilon^\prime,
S_v\cdot\epsilon]\{8[2ic_3^\prime(t_3)_{ti}-ic_4^\prime
(1-t_3)_{ti}]+{1\over
m_B}(1-t_3)_{ti}\}\nonumber\\
&&\times[J_2(x\omega,m^2_\pi)+J_2(-x\omega,m^2_\pi)]\nonumber\\
T_3d&=&i\omega C[S_v\cdot\epsilon^\prime,
S_v\cdot\epsilon]\{8i[c_3^\prime(t_3)_{ti}+c_4^\prime
Q_{ti}]-{1\over
m_B}Q_{ti}\}\{J_0(\omega,m^2_\pi)+J_0(-\omega,m^2_\pi)\}
\end{eqnarray}
where
\begin{eqnarray}
C=\frac{8ie^2D^2}{f^2_\pi}
\end{eqnarray}

The other diagrams (3e-3h) does not contribute to the spin
dependent Compton scattering.


\end {document}